\documentclass[twocolumn,preprintnumbers,amsmath,amssymb]{revtex4}

\usepackage{bm}
\usepackage{graphicx}
\usepackage{dcolumn}
\usepackage{color}

\begin{document}

\title{Lateral diffusion on a frozen random surface}
\author{Takao Ohta}
\email{takaoohta1949@gmail.com} 
\affiliation{Center for Integrative Medicine and Physics, Institute for Advanced Study, \\
Kyoto University, Kyoto 606-8302, Japan}
\author{Shigeyuki Komura }
\email{komura@tmu.ac.jp} 
\affiliation{Department of Chemistry, Graduate School of Science, Tokyo Metropolitan University, Tokyo 192-0397, Japan}


\begin{abstract}
The lateral diffusion coefficient of a Brownian particle on a two-dimensional random surface is studied in the quenched limit for which the surface configuration is time-independent. 
We start with the stochastic equation of motion for a Brownian particle on a fluctuating surface, which has been derived by Naji and Brown. The mean square displacement of the particle projected on 
a base 
plane is calculated exactly under the condition that  the surface with a constant shape has no spatial correlation. We prove that the obtained lateral diffusion coefficient  is in between the known upper and lower bounds.
\end{abstract}

\maketitle

{\it Introduction.}---Brownian motion on fluctuating surfaces has been of interest 
for
over the past four decades. The major reason is to elucidate the transport of biomolecules on cell membranes, which plays a crucial role for biological functions and 
signal 
processing in living cells~\cite{Marguet}. 
Advances in 
experimental techniques such as fluorescence photobleaching recovery~\cite{Axelrod,
Sprague} and single-particle tracking~\cite{
Lee, Kusumi} motivated theoretical studies of diffusion on
 a  restricted geometry. 
Safman and Delbr\"{u}ck introduced a model of Brownian motion in fluid membranes to calculate a translational mobility of a Brownian particle \cite{Safmann}. This approach has been developed, together with experimental studies, to describe the diffusion in flat membranes \cite{Block}. 

Lateral diffusion 
 on curved geometries
has been investigated theoretically both on 
static surfaces~\cite{Aizenbud, Halle} and dynamically fluctuating membranes~\cite{Halle, Gustafsson, Naji1, Naji2, Seifert}. The diffusion coefficient due to the geometrical effect of a curved surface and that arising from the interaction between the Brownian particle and membrane has been obtained. Most of the dynamical studies are concerned with the annealed case, in which motion of 
the
 Brownian particle is much slower than the shape relaxation of the membrane.
One of the present authors~\cite{Ohta} has formulated diffusion on a fluctuating surface starting with the bulk Brownian motion with 
the constraint that the particle is always on the surface.
It has been shown that the diffusion coefficient in the annealed case has a new contribution arising from the time-correlation of the lateral components of the surface velocity as well as the 
ordinary 
geometrical part~\cite{Ohta}.

\begin{figure}[tbh]
\begin{center}
\includegraphics[width=0.7\linewidth]{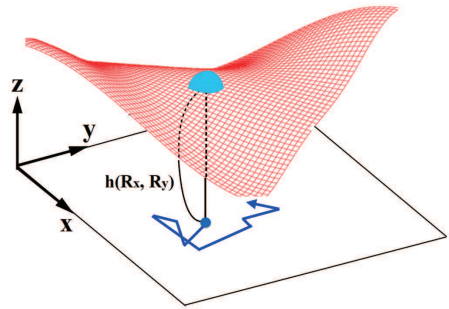}
\caption{(Color online) The height variable $h(x,y)$ for a random surface indicated by the red mesh.  The position of the Brownian particle (drawn in light blue) is given by $\vec{R}=(R_x, R_y, R_z)$ with $R_z=h(R_x, R_y)$. }
\label{Brown}
\end{center}
\end{figure}

The quenched case where the shape relaxation is slow compared with the motion of a Brownian particle has been investigated less intensively.
In one dimension, it has been shown that diffusion on a static surface is equivalent with Brownian motion in potential fields~\cite{Zwanzig, Halle}. The formula for the diffusion constant has been derived for both periodic~\cite{Lifson, Jackson, Festa} and random potentials~\cite{Zwanzig, Halle}. 
However, to the authors' knowledge, no exact theory for the diffusion coefficient has been available so far for two-dimensional surfaces. 
We address this problem in the present study. 

The quenched situation is expected in soft matter in thermal equilibrium, where interconnected structures with complex curved surfaces are often observed. Diffusion on such structures is highly non-trivial and the surface dynamics is not always relevant. 
Typical examples are organelles in living cells \cite{Sbalzarini} and the sponge phase in microemulsions \cite{Gelbart}. 
 Actually simulations of diffusion on static curved biological  surfaces have been conducted \cite{Sbalzarini}.

One may also expect the quenched situation in non-equilibrium systems. 
For example, the migrating velocity of an active Brownian particle which undergoes self-propulsion consuming energy 
produced inside or on the surface of the particle \cite{Ebeling, Ganguly} can be larger than the velocity caused only by thermal fluctuations. 
Quite recently, such studies in confined geometries have been carried out both experimentally \cite{Wang} and theoretically \cite{Castro}. 
When a membrane is subjected to non-thermal noise out of equilibrium \cite{Ramaswamy}, it might also increase the migration velocity. 

The application of the present theory is not limited to the systems mentioned above. It is possible, for example, that the result is useful for understanding chemical kinetics of adatoms on rough solid surfaces \cite{Masel}. 

{\it Langevin equation.}---We consider a Brownian particle migrating on a surface 
in a simple model system
as schematically  shown in Fig.~\ref{Brown}. The thickness of the surface is regarded as infinitesimal by assuming that the particle radius is much larger. 
Let us suppose that the particle is located at $\vec{R}(t)=(R_x(t), R_y(t), R_z(t))$ with $R_z=h(\vec{R}^{\perp}(t), t)$ and $\vec{R}^{\perp}=(R_x, R_y)$ where $h(\vec{r},t)$ is the height of the surface at $\vec{r}$ on the base plane. 
Naji and Brown have formulated a theory for Brownian motion on a fluctuating surface to derive the stochastic equation for the particle~\cite{Naji1}. The random force acting on the particle is defined on the tangent plane at each point on the surface. They solved numerically the Langevin equation coupled with the time-evolution equation for the height variable to investigate the crossover from the annealed to quenched cases.

The stochastic equation of motion for a Brownian particle in the Naji-Brown theory~\cite{Naji1} takes the form
 \begin{eqnarray}
\partial_t\vec{R}^{\perp}  = Z\vec{\nu}, 
  \label{eqRnu} 
 \end{eqnarray}
where $Z$ is a $2\times 2$ matrix whose  
components are given by 
\begin{eqnarray}  
 Z_{ij}
=\delta_{ij}-  \frac{(\partial_i h)(\partial_j h)}{g+\sqrt{g}}   
\label{Z} 
\end{eqnarray} 
with 
\begin{eqnarray}   
g=1+(\partial_x h)^2+ (\partial_y h)^2.
\label{g} 
\end{eqnarray} 
 The argument $\vec{r}$ in the height variable should be replaced by $\vec{R}^{\perp}$ after taking spatial derivative. 
 The random force $\vec{\nu}$ in Eq.~(\ref{eqRnu})
 obeys Gaussian statistics with zero mean and  
\begin{eqnarray}
\langle \nu_i(t) \nu_j(t')\rangle 
=2D_0\delta_{ij}\delta(t-t'). 
  \label{corr} 
 \end{eqnarray}  
The positive constant $D_0$ is the diffusion coefficient on a flat plane.  
 Since $Z$ depends on $\vec{\nu}$ through $h(\vec{R}^{\perp})$, the noise term in Eq.~(\ref{eqRnu}) is generally multiplicative. We employ the 
 Stratonovich interpretation of the multiplicative noise, while Naji and Brown take the Ito prescription with an additive drift term in Eq.~(\ref{eqRnu}). This difference is not essential for deriving the diffusion coefficient. 
 Naji and Brown have shown that~\cite{Naji1}
\begin{eqnarray}  
 Z_{ik}Z_{jk}=(g^{-1})_{ij},
 \label{ZZ} 
\end{eqnarray} 
 where  the repeated indices imply summation. 
The tensor  $ g^{-1}$  is the inverse of the metric tensor 
$g_{ij}=\delta_{ij}+(\partial_i h)(\partial_j h)$ and is given by
\begin{eqnarray}  
(g^{-1})_{ij}=\delta_{ij}-  \frac{(\partial_i h)(\partial_j h)}{g}.   
\label{ginverse} 
\end{eqnarray}

\begin{figure}[hbpt]
\begin{center}
\includegraphics[width=0.8\linewidth]{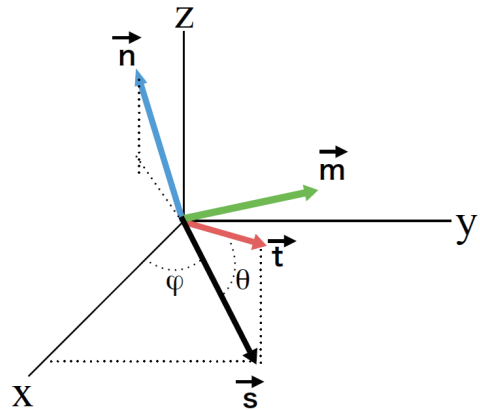}
\caption{(Color online) The blue vector $\vec{ n}$  is the normal unit vector, the red vector $\vec{t}$ the tangent unit vector, and the green vector $\vec{ m}$ the unit vector perpendicular to both $\vec{n}$ and $\vec{ t}$. The vector $\vec{s}$ ($|\vec{ s}|\ne 1$) is a projection of $\vec{ t}$ to the $x$-$y$ plane. The vector $\vec{m}$ is on the $x$-$y$ plane. The tangent plane touches the curved surface at $x=y=z=0$.}
\label{tangent plane}
\end{center}
\end{figure}

Here we describe an alternative derivation of the tensor $Z$, which has an intuitive implication. Let us introduce the tangent unit vector $\vec{t}$,  normal unit vector $\vec{n}$, and vector $\vec{m}$ which is orthogonal to the former two vectors as depicted in Fig.~\ref{tangent plane}. These unit vectors are defined, respectively, as
\begin{align}
\vec{t} &= \frac{1}{\sqrt{g}\sqrt{g-1}}(\partial_x h, \partial_y h, g-1)  \nonumber \\
&=(\cos\theta \cos\phi, \cos\theta \sin\phi, \sin\theta),
  \label{vect} \\
    \vec{n} &= \frac{1}{\sqrt{g}}(-\partial_x h, -\partial_y h, 1) \nonumber \\
&=(-\sin\theta\cos \phi, -\sin\theta\sin\phi, \cos\theta),
  \label{vecn} \\
\vec{m} &=-\frac{\vec{t}\times\vec{n}}{|\vec{t}\times\vec{n}|}= \frac{1}{\sqrt{g-1}}(-\partial_y h, \partial_x h, 0) \nonumber \\
&=(-\sin \phi, \cos\phi, 0),
  \label{vecm}  
   \end{align}
where $\theta$ is the angle between $\vec{t}$ and the $x$-$y$ plane and $\phi$ the angle between the projected vector $\vec{s}=(\partial_x h, \partial_y h, 0)/(\sqrt{g}\sqrt{g-1})$ and the $x$-axis as shown in Fig.~\ref{tangent plane} (notice that $|\vec{ s}|\ne 1$).
The unit vectors $\vec{t}$, $\vec{m}$ and $\vec{n}$ constitute  a set of orthogonal coordinates at each point on the curved surface.

The velocity of the Brownian particle on the left hand side of Eq.~(\ref{eqRnu}) is defined on the base plane, whereas the random force on the right hand side is parallel to the tangent plane~\cite{Naji1}. Keeping this fact in mind, we consider a mapping from the tangent to base planes. 
Let us introduce a vector on the tangent plane $\xi_t \vec{t}+\xi_m\vec{m}$ with arbitrary constants $\xi_t$ and $\xi_m$. The projection of this vector on the $x$-$y$ plane can be written as
\begin{eqnarray}
\eta_i = \xi_t t_i+ \xi_m m_i =Q_{ij}\xi_j,
  \label{eta} 
 \end{eqnarray}
where the third component is excluded, i.e., $i, j=1, 2$ and $\xi_1=\xi_t$ and $\xi_2=\xi_m$. 
From Eqs.~(\ref{vect}) and (\ref{vecm}), we find that the matrix  $Q$ takes the form 
\begin{eqnarray}
Q=
\left(
\begin{array} {ccc}
 \dfrac{\partial_x h}{\sqrt{g}\sqrt{g-1}}&  \dfrac{-\partial_y h}{\sqrt{g-1}} \\ 
\dfrac{\partial_y h}{\sqrt{g}\sqrt{g-1}}&\dfrac{\partial_x h}{\sqrt{g-1}}  
\end{array}  \right).
 \label{matrixP} 
 \end{eqnarray}
 However this is inadequate to enter directly  in the Langevin equation (\ref{eqRnu}) since 
 the off-diagonal elements of $Q$ are 
 not symmetric under the interchange $x\leftrightarrow y$, which is required by the isotropy of the base plane. 
We note that 
\begin{eqnarray}
P=QA
  \label{P} 
 \end{eqnarray}
 with
\begin{eqnarray}
A=\frac{1}{\sqrt{g-1}}\left(
\begin{array} {ccc}
\partial_x h&\partial_y h \\
-\partial_y h&\partial_x h 
\end{array}  \right)
=\left(
\begin{array} {ccc}
 \cos\phi&\sin\phi \\
-\sin\phi & \cos\phi 
\end{array}  \right)
 \label{A}
  \end{eqnarray}
is the only choice in terms of the angle $\phi$ to satisfy the symmetry requirement. In fact, 
it is readily proved from Eqs.~(\ref{matrixP}) and (\ref{A})  that $P$ is equal to $Z$ given 
by Eq.~(\ref{Z}). The relation in Eq.~(\ref{ZZ}) is also verified by noting that 
\begin{eqnarray}  
 Q_{ik}Q_{jk}=(g^{-1})_{ij}
  \label{QQ} 
 \end{eqnarray}
and $A_{ji}=(A^{-1})_{ij}$. 
The mapping in Eq.~(\ref{eta}) is then written as
\begin{eqnarray}
\vec{\eta} =P\vec{\zeta}
\label{eta2} 
\end{eqnarray}
with $\vec{\zeta}=A^{-1}\vec{\xi}$.

{\it Mean square displacement and lateral diffusion coefficient.}---Now we solve Eq.~(\ref{eqRnu}) in the case of a random time-independent surface. The solution can be written formally as
 \begin{align}
R_i^{\perp}(t)-R_i^{\perp}(0)  &= \int d\vec{r} \, Z(h(\vec{r}))_{ij}  \nonumber \\
&\times \int_0^t ds \, \delta(\vec{r}-\vec{R}^{\perp}(s))\nu_j(s),
  \label{eqRhat3} 
 \end{align}
from which one obtains
 \begin{align}
(\vec{R}^{\perp}(t)-\vec{R}^{\perp}(0))^2  &=\int d\vec{r}_1 \int  d\vec{r}_2 \, Z(h(\vec{r}_1))_{ij}Z(h(\vec{r}_2))_{ik} \nonumber \\
&\times\int_0^t ds_1\int_0^t ds_2 \, \delta(\vec{r}_1-\vec{R}^{\perp}(s_1))\nu_j(s_1) \nonumber \\
&\times\delta(\vec{r}_2-\vec{R}^{\perp}(s_2))\nu_k(s_2).
  \label{eqRhat4} 
 \end{align}

The mean square displacement on the base plane can be obtained by averaging Eq.~(\ref{eqRhat4}) over the randomness of the surface height and the random force. Since these are generally independent of each other for a frozen surface, one may write Eq.~(\ref{eqRhat4}) as
 \begin{align}
&\langle (\vec{R}^{\perp}(t)-\vec{R}^{\perp}(0))^2 \rangle  \nonumber \\
&= 
  \int d\vec{r}_1 \int d\vec{r}_2 \, \langle Z(h(\vec{r}_1))_{ij}Z(h(\vec{r}_2))_{ik}\rangle_h \nonumber \\
&\times \int_0^t ds_1\int_0^t ds_2 \, \langle \delta(\vec{r}_1-\vec{R}^{\perp}(s_1))\delta(\vec{r}_2-\vec{R}^{\perp}(s_2))\nonumber \\
&\times \nu_j(s_1)\nu_k(s_2) \rangle_{\vec{\nu}},
  \label{Gmsd} 
 \end{align}
where $\langle \cdots \rangle_h$ and $\langle \cdots \rangle_{\vec{\nu}}$ indicate the average over 
$h$ and $\vec{\nu}$, respectively.
Since we have assumed that the randomness of the height does not have any spatial correlation, the decoupling $\langle Z(h(\vec{r}_1))_{ij}Z(h(\vec{r}_2))_{ik}\rangle_h=\langle Z(h(\vec{r}_1))_{ij}\rangle_h\langle Z(h(\vec{r}_2))_{ik}\rangle_h$ is allowed and each part is constant in space because of the translational invariance. Therefore, one has
 \begin{eqnarray}
\langle (\vec{R}^{\perp}(t)-\vec{R}^{\perp}(0))^2 \rangle  =\langle Z_{ij}\rangle_h\langle Z_{ij}\rangle_h
2D_0 t=2dDt,
  \label{Gmsd2} 
 \end{eqnarray}
 where $d$ is the dimensionality of the surface and we have used Eq.~(\ref{corr}). The effective diffusion coefficient $D$ will be obtained shortly. 
The expression in Eq.~(\ref{Gmsd2}) implies that  taking an average of $Z$ in the Langevin equation (\ref{eqRnu}) with respect to the surface randomness is justified. 
By using  the isotropy of space, the effective lateral diffusion coefficient  $D$ is given for $d=2$ by
\begin{eqnarray}
\frac{D}{D_0}=\langle Z_{11}\rangle^2_h=\left \langle \frac{1}{2}\left(1+\frac{1}{\sqrt{g}}\right)\right \rangle_h^2.
\label{LD} 
\end{eqnarray}
This is the main result of this Letter.
When the surface is a curved line embedded in two dimensions, on the other hand, 
only the element $Z_{11}=1/\sqrt{g}$ 
exists
and one obtains for $d=1$
 \begin{eqnarray}
\frac{D}{D_0}=\left \langle \frac{1}{\sqrt{g}}\right \rangle_h ^2.
  \label{LD1} 
 \end{eqnarray}
This is the known exact result~\cite{Halle, Naji1} and has also been obtained by the present 
method~\cite{Ohta,comment}.

The ensemble average can be replaced by the surface average (the contour average in one dimension) defined 
 as~\cite{Gustafsson}
\begin{eqnarray}
\langle A\rangle_{s}  \equiv \frac{\int_0^L\int_0^Ldx \,dy\, \sqrt{g} A}{ \int_0^L\int_0^Ldx \, dy\, \sqrt{g}},
 \label{surf} 
\end{eqnarray}
where $L$ is the system size.
Hereafter we shall use this notation of the average $\langle \cdots \rangle_s$.

\begin{figure}[htb]
\begin{center}
\includegraphics[scale=0.35]{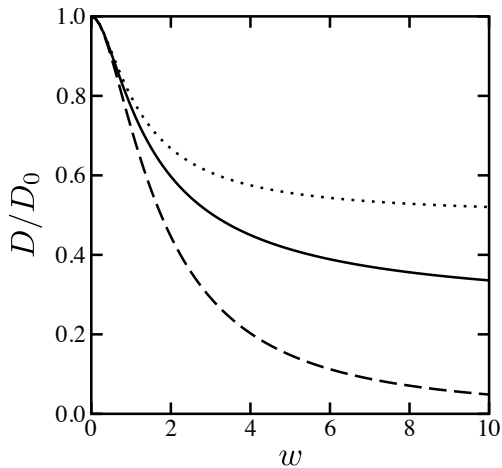}
\end{center}
\caption{
The scaled 
diffusion coefficient $D/D_0$ (solid line), the upper and lower bounds $D_{U}/D_0$ (dotted line), and $D_{L}/D_0$ (dashed line)
 given by Eqs.~(\ref{exact}),~(\ref{upper}), and (\ref{lower}), respectively, 
as a function of $w = \langle (\vec{\nabla}h)^{2} \rangle^{1/2}$ for a quenched 
Gaussian random surface.
}
\label{fig:diffusion}
\end{figure}

The upper and lower bounds for the diffusion coefficient in two dimensions have been obtained, respectively,  
as~\cite{Gustafsson}
\begin{equation}
\frac{D_U}{D_0}=  \frac{1}{2}\left(1+ \left\langle \frac{1}{g }\right\rangle_{s}\right), 
 \label{DU} 
\end{equation}
\begin{equation}
\frac{D_L}{D_0}= \frac{2 \left \langle 1/\sqrt{g} \right\rangle_s^2}{1+\left \langle 1/g \right\rangle_{s}}.
 \label{DL} 
\end{equation}
In order to prove $D_L \le D \le D_U$,  the inequality
\begin{equation}
 \langle A\rangle^2 \le \langle A^2\rangle
 \label{ineq} 
\end{equation}
is useful.
Applying this to Eq.~(\ref{LD}), one has
\begin{align}
\frac{D}{D_0 }& =\left(\left\langle  \frac{1}{2}\left(1+ \frac{1}{\sqrt{g}  }\right)\right\rangle_{s} \right)^2 
\nonumber \\
& \le \left\langle  \left( \frac{1}{2}\left(1+ \frac{1}{\sqrt{g}  }\right) \right)^2\right\rangle_{s}.
 \label{D2} 
\end{align}
It is readily shown that the last expression in Eq.~(\ref{D2}) is smaller than $D_U$ so that $D\le D_U$.

On the other hand, the lower bound Eq.~(\ref{DL}) satisfies
\begin{equation}
\frac{D_L}{D_0}=  \frac{2\left\langle 1/\sqrt{g} \right\rangle_s^2}{1+\left \langle 1/g \right \rangle_{s}} \le  \frac{2\left \langle 1/\sqrt{g} \right \rangle_s^2}{1+\left \langle 1/\sqrt{g} \right \rangle^2_{s}}.
 \label{DLL} 
\end{equation}
It is easy to see that $D$ is larger than the last expression of Eq.~(\ref{DLL}) leading to $D_L \le D$.

In the above, we have presented a mathematically rigorous proof for $D_L \le D \le D_U$. It might be useful, however, to show the relative magnitude and all over behavior of these three quantities. This is possible if the probability distribution of $\vec{p}=\vec{\nabla}h$ is Gaussian~\cite{Gustafsson}
\begin{equation}
f(p)=\frac{2}{w^{2}} \exp \Big(-\frac{\vec{p}^{2} }{w^{2}}\Big),
\label{gaussian}
\end{equation}
where $w = \langle (\vec{\nabla}h)^{2} \rangle^{1/2}$ is the root-mean-square fluctuation. The function $f(p)$ is normalized as
$\int_0^{\infty}dp\, p f(p)=1$.
The exact 
diffusion coefficient $D$ in Eq.~(\ref{LD}) turns out to be 
\begin{equation}
\frac{D}{D_{0}}=\frac{1}{4} [1+\sqrt{\pi} (1/w) \exp (1/w^{2}) \, {\rm erfc}(1/w)]^2,
\label{exact}
\end{equation}
where ${\rm erfc}(x)=(2/\pi)\int_x^{\infty} dt \, e^{-t^2}$.
On the other hand, the explicit forms of the upper and lower bounds given by Eqs.~(\ref{DU}) and (\ref{DL}), respectively, have been obtained 
for Gaussian surfaces as~\cite{Gustafsson}
\begin{equation}
\frac{D_{U}}{D_{0}} =\frac{1}{2} \left[ 1+ (1/w^{2}) \exp (1/w^{2}) E_{1}(1/w^{2}) \right], 
\label{upper}
\end{equation}
\begin{equation}
\frac{D_{L}}{D_{0}} = \frac{2\pi [(1/w)\exp (1/w^{2}) \, {\rm erfc}(1/w)]^2}
{1+(1/w^{2}) \exp (1/w^{2}) E_{1}(1/w^{2})},
\label{lower}
\end{equation}
where $E_1(x)=\int_x^{\infty} dt \, e^{-t}/t$.
In Fig.~\ref{fig:diffusion}, Eqs.~(\ref{exact}), (\ref{upper}), and (\ref{lower}) are plotted as a function of $w$. The diffusion coefficient $D/D_0$ decreases monotonically as $w$ is increased, approaching to the asymptotic value of 1/4.

{\it Discussion.}---Gustafsson and Halle~\cite{Gustafsson}  have argued that there are two candidates for the diffusion coefficient on the 
frozen disordered surface. Although our result in Eq.~(\ref{LD}) is exact, we compare it with other formulas. 
One is the result obtained by the area scaling in analogy with the projected contour length in one dimension~\cite{Halle, Gustafsson}
\begin{equation}
 \frac{D_A}{D_0} =\left \langle \frac{1}{\sqrt{g}} \right \rangle_s.
 \label{DA}
\end{equation}
The other is obtained by the effective medium approximation~\cite{Gustafsson}
\begin{equation}
 \frac{D_M}{D_0} =\frac{1}{\langle \sqrt{g}\rangle}_s.
 \label{DMD0}
\end{equation}
The requirement 
 $D_L\le D_A, D_M \le D_U$ has been verified for small values of $(\vec{\nabla}h)^2$~\cite{Gustafsson}.

One of the most distinct differences between the present result in Eq.~(\ref{LD}) and the approximants in  Eqs.~(\ref{DA}) and (\ref{DMD0}) is that $D$ is finite for $g\rightarrow \infty$, whereas both $D_A$ and $D_M$ vanish in this limit. Even if local maxima (minima) in a two-dimensional surface are extremely  high (deep), Brownian motion parallel to the base plane is always possible without being totally blocked so that it produces finite displacement. 
In fact, the elements of the matrix $Z$ for $\theta=\pi/2$ are given  by $Z_{11}=\sin^2\phi$, $Z_{12}=Z_{21}=-\sin\phi\cos\phi$ and $Z_{22}=\cos^2\phi$. The average over the angle $\phi$ yields $\langle Z_{11}\rangle_{\phi}\equiv (1/2\pi)\int_0^{2\pi}d\phi \, Z_{11}=1/2$, whose square 
gives rises to the factor $1/4$ in the diffusion coefficient $D$ in Eq.~(\ref{LD}). 
In contradiction to this, the 
diffusion coefficient obtained by numerical simulations of the Langevin equation seems consistent with $D_A$ even for large values of the surface roughness 
$w$
as shown in Figs.~9 and 10 in 
Ref.~\cite{Naji1}. The reason for this is unclear at present. 
It is geometrically obvious that the above argument does not hold in one dimension. Actually the diffusion constant in Eq.~(\ref{LD1}) vanishes in the limit $g\rightarrow \infty$

{\it Summary.}---In summary, the present theory provides an exact prediction for the diffusion coefficient on a two-dimensional frozen surface, which satisfies the restriction due to the upper and lower bounds. 
We start with the Langevin equation for a Brownian particle on a random surface formulated by Naji and Brown~\cite{Naji1}. The lateral diffusion coefficient is obtained by averaging the coefficient multiplied by the random force, which is valid in the quenched limit for random surfaces without any spatial correlation.

T.O.\ is indebted to Dr.\ Kazuhiko Seki for bringing his attention to Ref.~\cite{Zwanzig}.
S.K.\ thanks Yuto Hosaka for useful discussion.
S.K.\ acknowledges support by a Grant-in-Aid for Scientific Research (C) (Grant No.\ 18K03567 and
Grant No.\ 19K03765) from the Japan Society for the Promotion of Science 
and support by a Grant-in-Aid for Scientific Research on Innovative Areas
``Information Physics of Living Matters'' (Grant No.\ 20H05538) from the Ministry of Education, Culture, 
Sports, Science and Technology of Japan.


\begin{thebibliography}{99}
\bibitem{comment} 
The spatial average in Eq.~(A.13) in Ref.~\cite{Ohta} should read the surface (contour) average.
\end{thebibliography}
\end{document}